\RequirePackage{fix-cm}
\documentclass[twocolumn,epjc3]{svjour3}
\smartqed
\usepackage{amsmath,amssymb,graphicx}
\usepackage{placeins,booktabs,tabularx,float}
\usepackage[numbers,sort&compress]{natbib}
\usepackage{url}
\usepackage{microtype}
\AtBeginDocument{\thickmuskip=3mu plus 4mu\relax}
\AddToHook{cmd/bibliography/before}{\clearpage}

\journalname{Eur. Phys. J. C}
\begin{document}

\title{Polarization characteristics of synchrotron rings in Finslerian Kerr black holes and their dependence on emission conditions}

\titlerunning{Polarization characteristics of synchrotron rings in Finslerian Kerr black holes}
\author{Bohao Wang\thanksref{addr1,addr2}
  \and Hanyu Shi\thanksref{addr1}
  \and Xiaofeng Yang\thanksref{e1,addr1}
  \and Peiyang Li\thanksref{addr1}
  \and Yunliang Ren\thanksref{addr1,addr3}
  \and Xuwei Zhang\thanksref{addr1}
  \and Jianfeng Meng\thanksref{addr2,addr1}
  \and Jingze Li\thanksref{addr2,addr1}
  \and Xiongwei Liu\thanksref{addr2}}
\thankstext{e1}{e-mail: xfyang@henu.edu.cn}
\institute{School of Physics and Electronics, Henan University,
    No. 1 Jinming Street, Kaifeng, Henan 475004, China\label{addr1}
  \and School of Physics and Astronomy, China West Normal University,
    No. 1 Shida Road, Nanchong, Sichuan 637002, China\label{addr2}
  \and School of Physical Science and Technology, Xinjiang University,
    No. 777 Huarui Road, Urumqi, Xinjiang 830046, China\label{addr3}}
\date{\today}

\maketitle

\begin{abstract}
We investigate the polarized images of an equatorial
synchrotron-emitting ring in a Finslerian Kerr spacetime.
We trace light rays and parallel-transport the polarization to determine
how the deformation parameter changes the polarized intensity and the electric vector
position angle (EVPA) relative to their Kerr values. Among the five magnetic field
configurations examined at an inclination of $20^\circ$, the purely
vertical field produces the weakest polarized emission but the largest
local EVPA shift. For this field, increasing the inclination raises the
peak polarized intensity while reducing the maximum absolute EVPA shift.
Reversing the azimuthal flow shifts both the bright arc and the source
azimuth at which the largest EVPA shift occurs. Among the emission radii
examined, the largest local EVPA shift occurs near an intensity minimum
at $R=10M$. 
This local EVPA shift relative to the Kerr prediction may therefore be difficult to detect with current Event Horizon Telescope observations.
\end{abstract}

\bibliographystyle{spphys}

\section{Introduction}
\label{introduction}

The Event Horizon Telescope (EHT) Collaboration has released horizon-scale images of the supermassive black holes M87* and Sagittarius A* \cite{EHT2019I,EHT2019II,EHT2019III,EHT2019IV,EHT2019V,EHT2019VI,EHT2022SgrAI}, providing direct access to the emission regions in the immediate vicinity of black holes. 
The subsequent polarization images reveal spatially organized linear
polarization~\cite{EHT2021VII,EHT2024SgrAVII} and provide additional
information about the magnetic fields and plasma conditions near the
event horizon~\cite{EHT2021VIII,EHT2024SgrAVIII}. 
These observations motivate investigations of polarized synchrotron emission in strong gravitational fields. 

Detailed modeling of polarized black hole images often relies on numerical
simulations whose complexity and computational cost make broad parameter
surveys difficult. Narayan et al.\ developed a simplified
model for the polarized image of a synchrotron-emitting ring around a
Schwarzschild black hole~\cite{Narayan2021}. Gelles et al.\ extended this
model to equatorial emission in the Kerr spacetime~\cite{Gelles2021}.
The model utilizes an emitting ring with a fixed radius as the basic unit, 
providing a convenient approach for analyzing the impact of geometry and launch conditions on polarization characteristics.
In the optically thin limit, the contributions to the Stokes parameters
from rings at different radii can be weighted by the radial emission
profile and summed to construct an image of a ring of finite
width~\cite{Gelles2021}.
More recently, polarized images of black holes have been further investigated \cite{Qin2022GB,Qin2022STVG,Liu2022Regular,ShiZhu2024,Chen2025EMS}.
This modeling framework has also been extended to spatially extended disks, non-equatorial emission geometries, and orbiting or inspiraling hotspots, allowing the effects of emission geometry and plasma motion on polarization signatures to be explored  \cite{Zhang2022Bonnor,Bernshteyn2026Cone,Tlemissov2026Hotspot,Ruales2026Inspiral}.In particular, Qin et al.\ examined how the deformation parameter of a
rotating Konoplya--Zhidenko black hole affects the polarized intensity
and electric vector position angle of an equatorial emitting
ring~\cite{Qin2026KZ}.

The standard cosmology based on general relativity has been extensively validated by observational experiments. However, due to certain unexplained physical effects, physicists have been compelled to postulate the existence of dark sectors.
These issues pose certain challenges to general relativity, prompting physicists to propose various alternative theories of gravity to address them. 
Recently, 
Nekouee proposed a Finslerian extension of the Kerr black hole within the Barthel–Randers framework \cite{Nekouee2026}.
This model introduces geometric deformation via an additional 1-form and employs the osculating Riemannian method to construct an effective spacetime.
The original study analyzed the impact of deformation on photon orbits and the innermost stable circular orbit.
Nekouee also found that, within a small deformation range of \(|\widetilde{\xi}|\lesssim 0.1\), the Finsler deformation can induce shifts in the characteristic orbital radii similar to those produced by changing the Kerr spin.

In this paper, we investigate the direct polarized images of an equatorial synchrotron emitting ring in the Finslerian Kerr effective spacetime. 
In the geometric-optics limit of Maxwell theory, light rays follow null
geodesics and the polarization vector is parallel-transported along each
ray~\cite{Dolan2018}. 
We apply this description to the effective metric
and integrate the corresponding equations numerically.
Taking the corresponding Kerr configuration as a reference, we examine the effects of the deformation parameter on the image-ring morphology, polarized intensity, and $\mathrm{EVPA}$. 
We consider three pure and two mixed magnetic field configurations. The four flow states are rest, prograde azimuthal flow, retrograde azimuthal flow, and radial inflow. 
We also investigate the dependence on spin and observer inclination, and compare rings at seven emission radii $R/M = 6, 8, 10, 12, 14, 16,$ and $1000$.

\section{Polarization and Observation of Black Hole Emission Ring in Finslerian Kerr Spacetime}

We adopt the Finslerian Kerr black hole solution introduced by Nekouee.
We use metric signature $(-,+,+,+)$ and set $G=c=M=1$,
where $M$ is the metric mass parameter.
Lengths are measured in units of $GM/c^2$.
We define the dimensionless deformation parameter as
$\widetilde{\xi}\equiv\xi/M^2$,
The radial coordinate $r$ and the spin parameter $a$ are expressed in units of $M$.
We consider a small deformation range of \(\widetilde{\xi} \in [-0.1, 0.1]\).

For a constant $\widetilde{\xi}$, the effective osculating
metric in Boyer--Lindquist coordinates is
\begin{equation}
\begin{aligned}
d\bar s^2
&=(1+\widetilde{\xi})
  \left(ds_{\rm K}^2+\widetilde{\xi}\,dt^2\right),\\
\bar g_{\mu\nu}
&=(1+\widetilde{\xi})
  \left(g^{\rm K}_{\mu\nu}
  +\widetilde{\xi}\,\delta_\mu^t\delta_\nu^t\right).
\end{aligned}
\label{eq:effective-metric}
\end{equation}
Here $ds_{\rm K}^2$ denotes the Kerr line element,
which takes the form
\begin{equation}
\begin{aligned}
ds_{\mathrm K}^{2}={}&
-\left(1-\frac{2r}{\Sigma}\right)dt^{2}
-\frac{4ar\sin^{2}\theta}{\Sigma}\,dt\,d\phi \\
&+\frac{\Sigma}{\Delta}\,dr^{2}
+\Sigma\,d\theta^{2} \\
&+\left(r^{2}+a^{2}
+\frac{2a^{2}r\sin^{2}\theta}{\Sigma}\right)
\sin^{2}\theta\,d\phi^{2},
\end{aligned}
\label{eq:kerr-line-element}
\end{equation}
where
\begin{equation}
\Sigma=r^{2}+a^{2}\cos^{2}\theta,
\qquad
\Delta=r^{2}-2r+a^{2}.
\label{eq:sigma-delta}
\end{equation}
Equation~\eqref{eq:effective-metric} recovers to the Kerr metric for $\widetilde{\xi}=0$ 
and to the Schwarzschild metric for $a=\widetilde{\xi}=0$.
The asymptotically normalized time and radial coordinates and the mass ratio are given by
\begin{equation}
 t'=\sqrt{1-\widetilde{\xi}^2}\,t,\quad
 r'=\sqrt{1+\widetilde{\xi}}\,r,\quad
 \frac{M_{\rm ADM}}{M}=\frac{\sqrt{1+\widetilde{\xi}}}{1-\widetilde{\xi}}.
 \label{eq:asymptotic}
\end{equation}

\begin{figure}[htb]
	\centering
	\includegraphics[width=\linewidth]{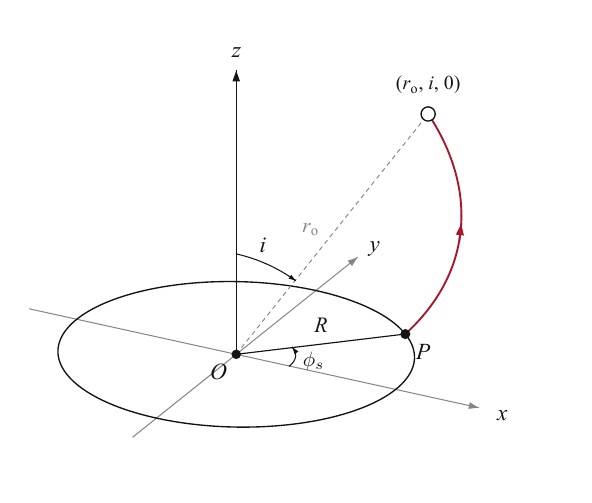}
    \caption{Geometry of the equatorial emitting ring. A point
\(P\) on the ring has spatial coordinates
\((r,\theta,\phi)=(R,\pi/2,\phi_s)\), where \(\phi_s\) is the source
azimuth. The observer is located at \((r_o,i,0)\), with \(r_o=10^5M\)
and inclination \(i\) measured from the rotation axis.}
    \label{fig:emission-geometry}
\end{figure}

We assume minimal coupling between the simple Maxwell field and the effective metric $\bar g_{\mu\nu}$.
As shown in Fig.~\ref{fig:emission-geometry}, we consider a thin emitting
ring of radius \(R\) in the equatorial plane of the black hole. 

In the propagation region where plasma refraction is neglected,
photons satisfy the null geodesic equation, and polarization satisfies the parallel transport equation.
To simplify the integral computation,
we use the Hamiltonian:
\begin{equation}
 \mathcal H=\frac12\bar g^{\mu\nu}p_\mu p_\nu=0.
 \label{eq:hamiltonian}
\end{equation}
Here, $k^\mu$ denotes the future-directed photon
wave four-vector along the null ray connecting the emitter
to the observer, and
$p_\mu=\bar g_{\mu\nu}k^\nu$ is its covariant counterpart.
The ray trajectory obeys Hamilton’s equations,
\begin{align}
 \frac{dx^\mu}{d\lambda}&=\bar g^{\mu\nu}p_\nu,
 \label{eq:ray-position}\\
 \frac{dp_\mu}{d\lambda}
 &=-\frac12\partial_\mu\bar g^{\alpha\beta}p_\alpha p_\beta
 =\frac12\partial_\mu\bar g_{\alpha\beta}k^\alpha k^\beta,
 \label{eq:ray-momentum}
\end{align}
where $\lambda$ is the affine parameter.
Due to the stationary background,
the corresponding conserved quantities along each null ray are 
\begin{equation} 
E \equiv -p_t = \mathrm{const.}, 
\qquad p_\phi = \mathrm{const.} 
\label{eq:photon_constants} 
\end{equation}

\begin{figure}[htb]
	\centering
	\includegraphics[width=\linewidth]{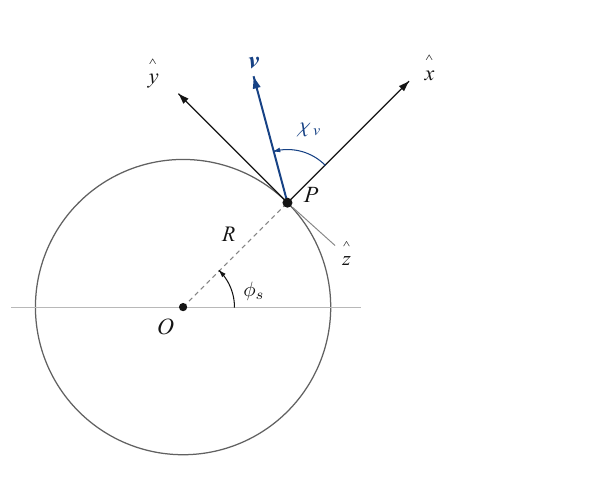}
	\caption{Local Z-frame at the emission point \(P\). The blue arrow
shows the fluid velocity \(\mathbf{v}\), and the blue arc marks its
angle \(\chi_v\) from the outward radial axis.}
    \label{fig:z-frame}
\end{figure}

To describe the photon momentum and the fluid velocity in a
local orthonormal frame,  we introduce a zero-angular-momentum observer
(ZAMO) at the emission point $P$ shown in Fig.~\ref{fig:z-frame}. 

The observer remains at fixed $r$ and $\theta$, while moving
in the azimuthal direction with coordinate angular velocity
$\omega=d\phi/dt$. The condition of vanishing angular momentum,
$u_{{\rm Z}\phi}=0$, gives
\begin{equation}
    \omega=-\frac{\bar g_{t\phi}}{\bar g_{\phi\phi}}.
    \label{eq:zamo_omega}
\end{equation}
The normalization
$\bar g_{\mu\nu}u_{\rm Z}^{\mu}u_{\rm Z}^{\nu}=-1$
then determines the lapse function,
\begin{equation}
    N=\sqrt{
        -\bar g_{tt}
        +\frac{\bar g_{t\phi}^{\,2}}{\bar g_{\phi\phi}}
    },
    \label{eq:zamo_lapse}
\end{equation}
so that $d\tau_{\rm Z}=N\,dt$, where $\tau_{\rm Z}$ is the
proper time of the ZAMO. 

We refer to the local orthonormal frame carried by this
observer as the Z-frame. Its spatial axes
$(\hat{\boldsymbol{x}},\hat{\boldsymbol{y}},
\hat{\boldsymbol{z}})$ point in the outward radial direction,
the direction of increasing $\phi$, and the direction normal
to the equatorial plane toward decreasing $\theta$,
respectively. The corresponding orthonormal tetrad is
\begin{equation}
\begin{aligned}
    e_{(0)}=u_{\rm Z}
    &=\frac{1}{N}
      \left(\partial_t+\omega\partial_\phi\right),
    &
    e_{(x)}
    &=\frac{1}{\sqrt{\bar g_{rr}}}\,\partial_r,
    \\[4pt]
    e_{(y)}
    &=\frac{1}{\sqrt{\bar g_{\phi\phi}}}\,\partial_\phi,
    &
    e_{(z)}
    &=-\frac{1}{\sqrt{\bar g_{\theta\theta}}}\,\partial_\theta,
\end{aligned}
\label{eq:zamo_tetrad}
\end{equation}
where $\partial_\mu$ denotes the coordinate basis vector.
At any emission point, 
the Z-frame components of the photon wave four-vector are obtained by projecting its covariant momentum onto the tetrad:
\begin{align}
 k_{(Z)}^{\hat t}&=-p_\mu e_{(0)}^\mu
 =\frac{E-\omega p_\phi}{N},
 \label{eq:local-kt}\\
 k_{(Z)}^{\hat x}&=\frac{p_r}{\sqrt{\bar g_{rr}}},
 \label{eq:local-kx}\\
 k_{(Z)}^{\hat y}&=\frac{p_\phi}{\sqrt{\bar g_{\phi\phi}}},
 \label{eq:local-ky}\\
 k_{(Z)}^{\hat z}&=-\frac{p_\theta}{\sqrt{\bar g_{\theta\theta}}}.
 \label{eq:local-kz}
\end{align}

In the Z-frame, the fluid moves in the equatorial plane at speed $v$.
Its velocity makes an angle $\chi_v$ with the outward radial direction (Fig.~\ref{fig:z-frame}).
The velocity vector is therefore
\begin{equation}
 \mathbf v
 =v\cos\chi_v\,\hat x+v\sin\chi_v\,\hat y 
 \label{eq:velocity}
\end{equation}
The 4-velocity of the fluid is
\begin{equation}
 u_{\rm F}
 =\gamma\left(e_{(0)}
 +v\cos\chi_v\,e_{(x)}+v\sin\chi_v\,e_{(y)}\right),
 \label{eq:fluid-four-velocity}
\end{equation}
where $\gamma=(1-v^2)^{-1/2}$.

Applying a Lorentz boost from the Z-frame to the fluid rest frame (the F-frame), we obtain
\begin{align}
 k_{(F)}^{\hat t}
 &=\gamma k_{(Z)}^{\hat t}
 -\gamma v\cos\chi_v\,k_{(Z)}^{\hat x}
 -\gamma v\sin\chi_v\,k_{(Z)}^{\hat y},
 \label{eq:boost-kt}\\
 k_{(F)}^{\hat x}
 &=-\gamma v\cos\chi_v\,k_{(Z)}^{\hat t}
 +\left[1+(\gamma-1)\cos^2\chi_v\right]k_{(Z)}^{\hat x}
 \nonumber\\[-0.2em]
 &\quad+(\gamma-1)\cos\chi_v\sin\chi_v\,k_{(Z)}^{\hat y},
 \label{eq:boost-kx}\\
 k_{(F)}^{\hat y}
 &=-\gamma v\sin\chi_v\,k_{(Z)}^{\hat t}
 +(\gamma-1)\cos\chi_v\sin\chi_v\,k_{(Z)}^{\hat x}
 \nonumber\\[-0.2em]
 &\quad+\left[1+(\gamma-1)\sin^2\chi_v\right]k_{(Z)}^{\hat y},
 \label{eq:boost-ky}\\
 k_{(F)}^{\hat z}&=k_{(Z)}^{\hat z}.
 \label{eq:boost-kz}
\end{align}
We define the unit photon propagation vector in the fluid frame by $\mathbf n_{\rm F}=\mathbf k_{(F)}/|\mathbf k_{(F)}|$.
The observer’s four-velocity is $u_o^\mu=e_{(0)}^\mu|_o$.
The frequency-shift factor is then
\begin{equation}
 g_\nu=\frac{\nu_o}{\nu_{\rm F}}
 =\frac{-p_\mu u_o^\mu}{-p_\mu u_{\rm F}^\mu}
 =\frac{E_o}{k_{(F)}^{\hat t}},
 \qquad
 E_o=\left.\frac{E-\omega p_\phi}{N}\right|_o.
 \label{eq:redshift}
\end{equation}
The factor $g_\nu$ accounts for both the gravitational redshift and the Doppler effect due to fluid motion.

In the F-frame, the magnetic field is specified by its components
along the local radial, azimuthal, and vertical directions,
\begin{equation}
    \boldsymbol{B}
    = B_r \hat{\boldsymbol{x}}
    + B_\phi \hat{\boldsymbol{y}}
    + B_z \hat{\boldsymbol{z}}.
    \label{eq:magnetic_field}
\end{equation}
To isolate the effect of magnetic field orientation on the polarized
image, we use the same field strength
\(B_0=\sqrt{B_r^2+B_\phi^2+B_z^2}\) .
After normalizing each component by \(B_0\), we consider the following five magnetic field configurations
\begin{equation}
  \frac{(B_r,B_\phi,B_z)}{B_0}
  \in
  \left\{
  \begin{aligned}
    &(1,0,0),\ (0,1,0),\ (0,0,1),\\[2pt]
    &\frac{(1,1,0)}{\sqrt{2}},\quad
     \frac{(0,1,1)}{\sqrt{2}}
  \end{aligned}
  \right\}.
  \label{eq:magnetic_configurations}
\end{equation}
The first three configurations represent purely radial,
toroidal, and vertical magnetic fields, respectively.
The last two represent radial--toroidal and toroidal--vertical
mixed fields with equal component magnitudes.

The angle $\zeta$ between the photon direction and the magnetic field is given by
\begin{equation}
 \sin\zeta
 =\frac{|\mathbf k_{(F)}\times\mathbf B|}
 {|\mathbf k_{(F)}|\,B}
 =\frac{|\mathbf n_{\rm F}\times\mathbf B|}{B}.
 \label{eq:pitch-angle}
\end{equation}
For optically thin synchrotron radiation, 
the linear polarization electric vector is perpendicular to the projected magnetic field.
The unit polarization four-vector at the source can be written as
\begin{align}
 f_{(F)}^{\hat t}&=0,
 \label{eq:emission-ft}\\
 f_{(F)}^{\hat x}
 &=\frac{n_{\rm F}^{y}B_z-n_{\rm F}^{z}B_\phi}{B\sin\zeta},
 \label{eq:emission-fx}\\
 f_{(F)}^{\hat y}
 &=\frac{n_{\rm F}^{z}B_r-n_{\rm F}^{x}B_z}{B\sin\zeta},
 \label{eq:emission-fy}\\
 f_{(F)}^{\hat z}
 &=\frac{n_{\rm F}^{x}B_\phi-n_{\rm F}^{y}B_r}{B\sin\zeta}.
 \label{eq:emission-fz}
\end{align}
Thus, we obtain
\begin{equation}
 f^\mu k_\mu=0,\qquad f^\mu f_\mu=1.
 \label{eq:polarization-constraints}
\end{equation}
Applying the inverse Lorentz transformation, 
we obtain the polarization components in the Z-frame.
\begin{align}
 f_{(Z)}^{\hat t}
 &=\gamma f_{(F)}^{\hat t}
 +\gamma v\cos\chi_v\,f_{(F)}^{\hat x}
 +\gamma v\sin\chi_v\,f_{(F)}^{\hat y},
 \label{eq:inverse-ft}\\
 f_{(Z)}^{\hat x}
 &=\gamma v\cos\chi_v\,f_{(F)}^{\hat t}
 +\left[1+(\gamma-1)\cos^2\chi_v\right]f_{(F)}^{\hat x}
 \nonumber\\[-0.2em]
 &\quad+(\gamma-1)\cos\chi_v\sin\chi_v\,f_{(F)}^{\hat y},
 \label{eq:inverse-fx}\\
 f_{(Z)}^{\hat y}
 &=\gamma v\sin\chi_v\,f_{(F)}^{\hat t}
 +(\gamma-1)\cos\chi_v\sin\chi_v\,f_{(F)}^{\hat x}
 \nonumber\\[-0.2em]
 &\quad+\left[1+(\gamma-1)\sin^2\chi_v\right]f_{(F)}^{\hat y},
 \label{eq:inverse-fy}\\
 f_{(Z)}^{\hat z}&=f_{(F)}^{\hat z}.
 \label{eq:inverse-fz}
\end{align}
Using Eq.~\eqref{eq:zamo_tetrad}, we obtain the coordinate components of the polarization four-vector
\begin{align}
 f^t&=\frac{f_{(Z)}^{\hat t}}{N},
 &f^r&=\frac{f_{(Z)}^{\hat x}}{\sqrt{\bar g_{rr}}},
 \label{eq:coordinate-fr}\\
 f^\theta&=-\frac{f_{(Z)}^{\hat z}}{\sqrt{\bar g_{\theta\theta}}},
 &f^\phi&=\frac{\omega f_{(Z)}^{\hat t}}{N}
 +\frac{f_{(Z)}^{\hat y}}{\sqrt{\bar g_{\phi\phi}}}.
 \label{eq:coordinate-fa}
\end{align}
The time term in the azimuthal component arises from the local frame of the rotating background.
We define the horizontal and vertical screen directions using the observer’s local $+\hat y$ and $+\hat z$ axes, respectively. The unit spatial vector pointing from the observer toward the emitting region is then
\begin{equation}
 \mathbf n_{\rm back}
 =\frac{(-1,\alpha/D,\beta/D)}
 {\sqrt{1+(\alpha/D)^2+(\beta/D)^2}},
 \qquad
 D=\sqrt{1+\widetilde{\xi}}\,r_o.
 \label{eq:camera-direction}
\end{equation}
The corresponding screen coordinates are then given by
\begin{equation}
 \alpha=D\frac{n_{\rm back}^{y}}{-n_{\rm back}^{x}},
 \qquad
 \beta=D\frac{n_{\rm back}^{z}}{-n_{\rm back}^{x}}.
 \label{eq:screen-coordinates}
\end{equation}
Here $D$ is the asymptotic length scale adopted by the camera, and $\alpha/M$ and $\beta/M$ are the coordinates shown in the result figures. 
We construct the direct image of the emitting ring by tracing
null rays backward from the observer and selecting those whose
first equatorial-plane crossing occurs at $r=R$.
The past-directed photon momentum used in the backward
integration is reversed at the emission point when evaluating
the local emission quantities.

We next compute the linearly polarized specific intensity arriving at the observer along this ray.
In the F-frame, we assume that the emitting electrons follow an isotropic power-law energy distribution $N_e(\gamma_e)=K\gamma_e^{-p_e}$.
In the power-law regime, the emissivity is
$j_{\nu,{\rm F}}\propto(B\sin\zeta)^{1+s}\nu_{\rm F}^{-s}$,
where $s=(p_e-1)/2$ is the spectral index \cite{Pandya2016}.
Combining the invariance of $I_\nu/\nu^3$ in vacuum with the local optically thin approximation, we obtain the observed linearly polarized specific intensity
\begin{equation}
 I_p=g_\nu^{3+s}\ell_p B^{1+s}\sin^{1+s}\zeta.
 \label{eq:general-intensity}
\end{equation}
Here, $\ell_p$ denotes the path length of the photon through the emitting region.
Approximating the emitting region as a locally planar layer of proper thickness $H$, we evaluate this length as
\begin{equation}
 \ell_p=\frac{H}{|n_{\rm F}^{z}|}
 =\frac{Hk_{(F)}^{\hat t}}{|k_{(F)}^{\hat z}|},
 \label{eq:path-length}
\end{equation}
With the same normalization for all configurations and with $s=1$,
Equation \eqref{eq:general-intensity} becomes
\begin{equation}
 I_p=g_\nu^4\ell_p B^2\sin^2\zeta.
 \label{eq:intensity}
\end{equation}

The radiatively weighted components of the observed polarization amplitude are
\begin{align}
 \mathcal E_{\alpha,\rm obs}
 &=\sqrt{I_p}\,e_\alpha
 =\,g_\nu^2\ell_p^{1/2}B\sin\zeta\,e_\alpha,
 \label{eq:weighted-ealpha}\\
 \mathcal E_{\beta,\rm obs}
 &=\sqrt{I_p}\,e_\beta
 =\,g_\nu^2\ell_p^{1/2}B\sin\zeta\,e_\beta,
 \label{eq:weighted-ebeta}\\
 I_p&=\mathcal E_{\alpha,\rm obs}^2+\mathcal E_{\beta,\rm obs}^2.
 \label{eq:amplitude-intensity}
\end{align}
We measure the EVPA counterclockwise from the positive
$\beta$-axis on the observer's screen. 
The Stokes parameters and the EVPA are
\begin{align}
    Q &= \mathcal{E}_{\beta,\mathrm{obs}}^{2}
       - \mathcal{E}_{\alpha,\mathrm{obs}}^{2},
    \label{eq:stokes_q}\\
    U &= -2\mathcal{E}_{\alpha,\mathrm{obs}}
            \mathcal{E}_{\beta,\mathrm{obs}},
    \label{eq:stokes_u}\\
    \mathrm{EVPA}
      &= \frac{1}{2}\operatorname{atan2}(U,Q)
         \pmod{\pi}.
    \label{eq:evpa}
\end{align}

These quantities determine the polarized intensity and
polarization direction of the emitting ring.
In the polarization ring diagrams, the orientation of each
short bar represents the EVPA, while its length is proportional
to $I_p/I_{\mathrm{ref}}$.

To compare the deformed background with the Kerr reference,
we define the complex linear polarization as $P=Q+iU$.
The shortest signed EVPA difference at the same source
azimuth $\phi$ is then
\begin{equation}
    \Delta\mathrm{EVPA}(\phi)
    =\frac{1}{2}\operatorname{Arg}
    \left[P_{\widetilde{\xi}}(\phi)P_0^{*}(\phi)\right],
    \label{eq:delta_evpa}
\end{equation}
where $P_{\widetilde{\xi}}$ and $P_0$ denote the complex
polarization in the deformed and Kerr backgrounds,
respectively, and the asterisk denotes complex conjugation.
When plotting \(\Delta\mathrm{EVPA}\), we retain only those source
azimuths at which the polarized intensity in each model exceeds
\(1\%\) of that model's peak value. The retained azimuths form the set
\begin{equation}
 \mathcal S_\tau=
 \left\{\phi:
 I_{p,\widetilde{\xi}}(\phi)>\tau I_{p,\widetilde{\xi}}^{\max},\
 I_{p,0}(\phi)>\tau I_{p,0}^{\max}
 \right\},\tau=0.01.
 \label{eq:bright-mask}
\end{equation}

\section{Results}
We compute the polarized images of a synchrotron-emitting ring around
the Finslerian Kerr black hole. 
We use
\(\widetilde{\xi}\in\{-0.1,-0.05,0,+0.05,+0.1\}\) to investigate the
effects of the deformation on the polarized image. We vary the black hole
spin, fluid velocity, magnetic field configuration, emission radius, and
observer inclination to examine their effects on the ring morphology,
polarized intensity, and EVPA. 
Figures~\ref{spin} --~\ref{ER2}
are polarization images,and Figures~\ref{mag_fu_QU} and ~\ref{ERQU}
are Q-U diagrams.
\subsection{Effects of parameters on the polarization images}
\begin{figure*}[!htbp]
	\centering
	\includegraphics[width=\linewidth]{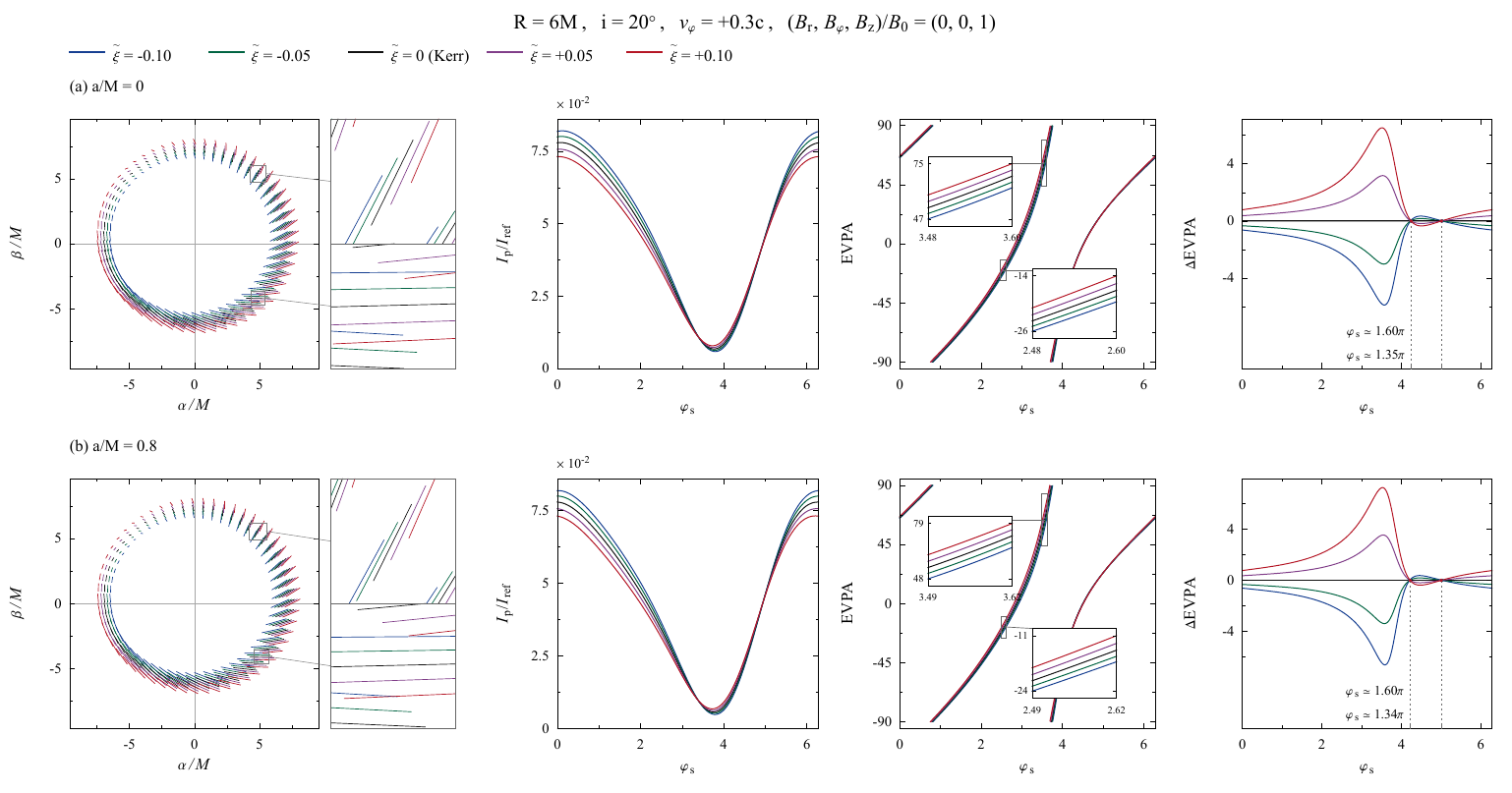}
	\caption{Spin comparison for a vertical field at \(R=6M\), \(i=20^\circ\), and \(v_\phi=+0.3c\). Rows (a) and (b) show \(a/M=0\) and \(0.8\), respectively. Bar orientation gives the EVPA and bar length is proportional to \(I_p/I_{\rm ref}\). Dotted vertical lines in the last column mark approximate zero-crossing azimuths.}
    \label{spin}
\end{figure*}

\begin{figure*}[!htbp]
	\centering
	\includegraphics[width=\linewidth]{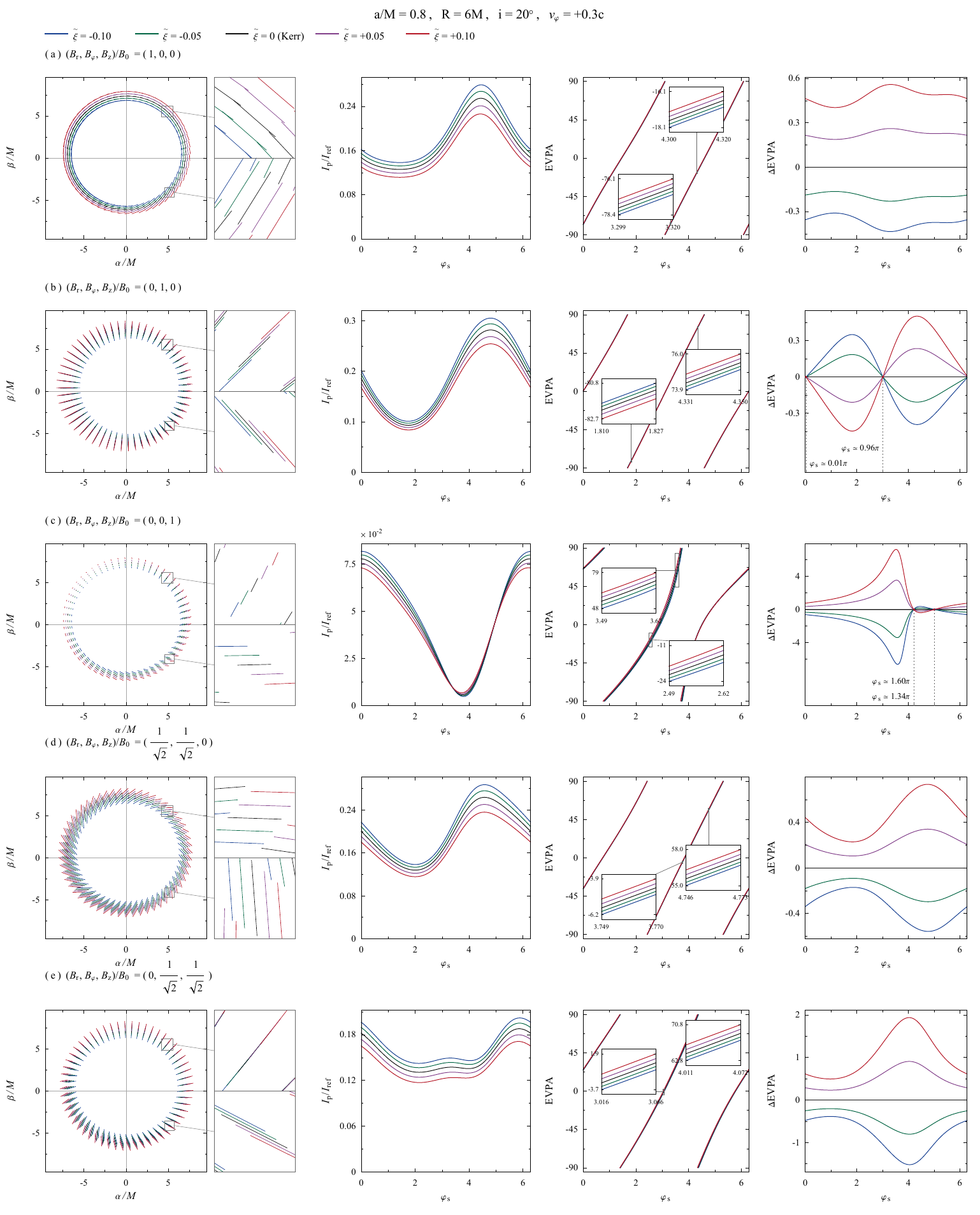}
    \caption{Five magnetic field orientations at fixed field strength, with \(a/M=0.8\), \(R=6M\), \(i=20^\circ\), and \(v_\phi=+0.3c\). Rows (a)--(c) show purely radial, toroidal, and vertical fields. Rows (d) and (e) combine equal radial and toroidal components or equal toroidal and vertical components, respectively.}
    \label{magnetic}
\end{figure*}

\begin{figure*}[!htbp]
	\centering
	\includegraphics[width=\linewidth]{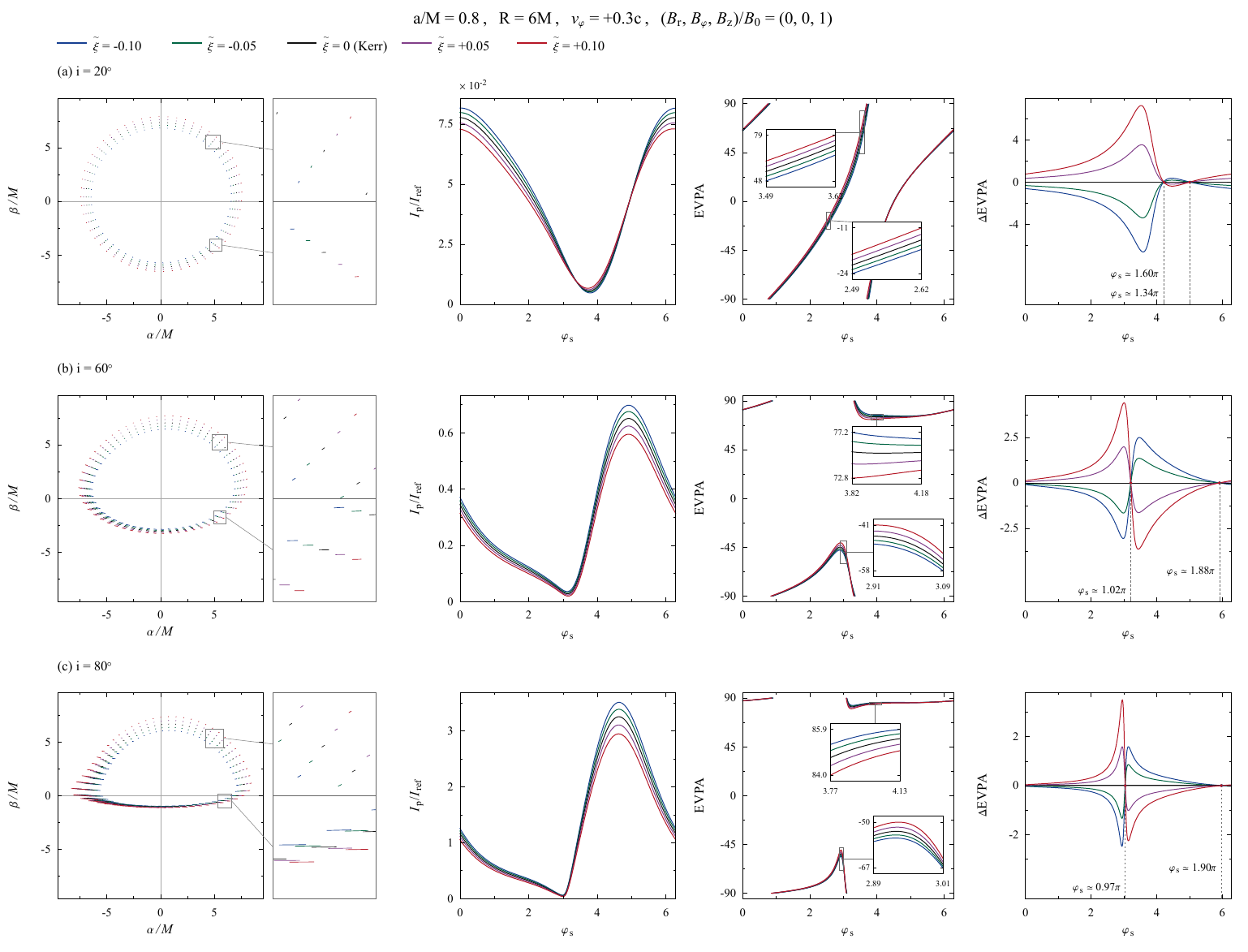}
	\caption{Inclination dependence for a vertical field at \(a/M=0.8\), \(R=6M\), and \(v_\phi=+0.3c\). Rows (a)--(c) follow the same ring from a nearly face-on view (\(20^\circ\)) to \(60^\circ\) and \(80^\circ\). }
    \label{inclination}
\end{figure*}

\begin{figure*}[!htbp]
	\centering
	\includegraphics[width=\linewidth]{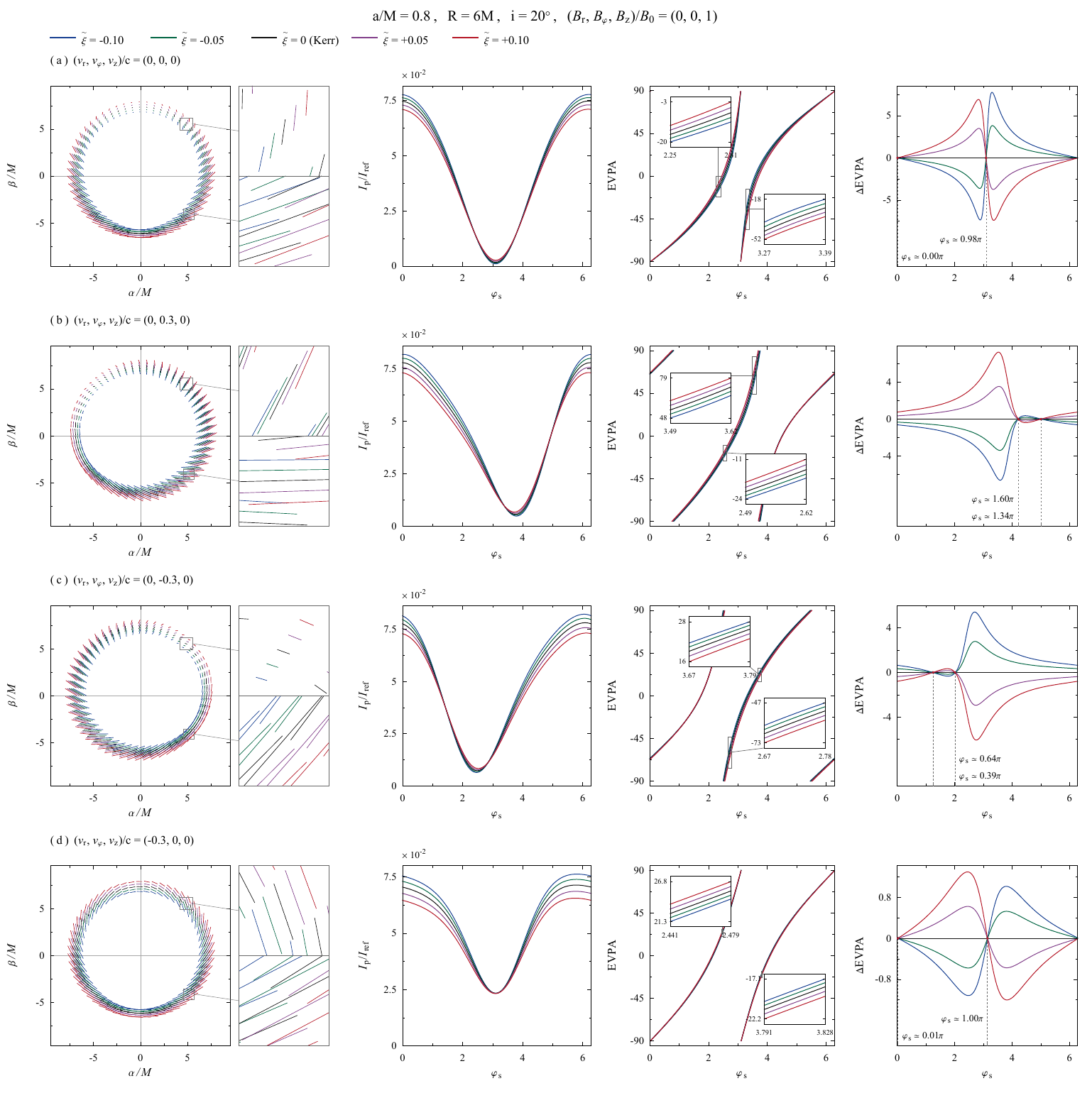}
	\caption{Four fluid velocities in a vertical magnetic field at \(a/M=0.8\), \(R=6M\), and \(i=20^\circ\). In row (a) the fluid is at rest relative to the local zero-angular-momentum observer. Rows (b)--(d) show prograde azimuthal flow, retrograde azimuthal flow, and radial inflow, all at \(0.3c\). }
    \label{fluid}
\end{figure*}

\begin{figure*}[!htbp]
	\centering
	\includegraphics[width=\linewidth]{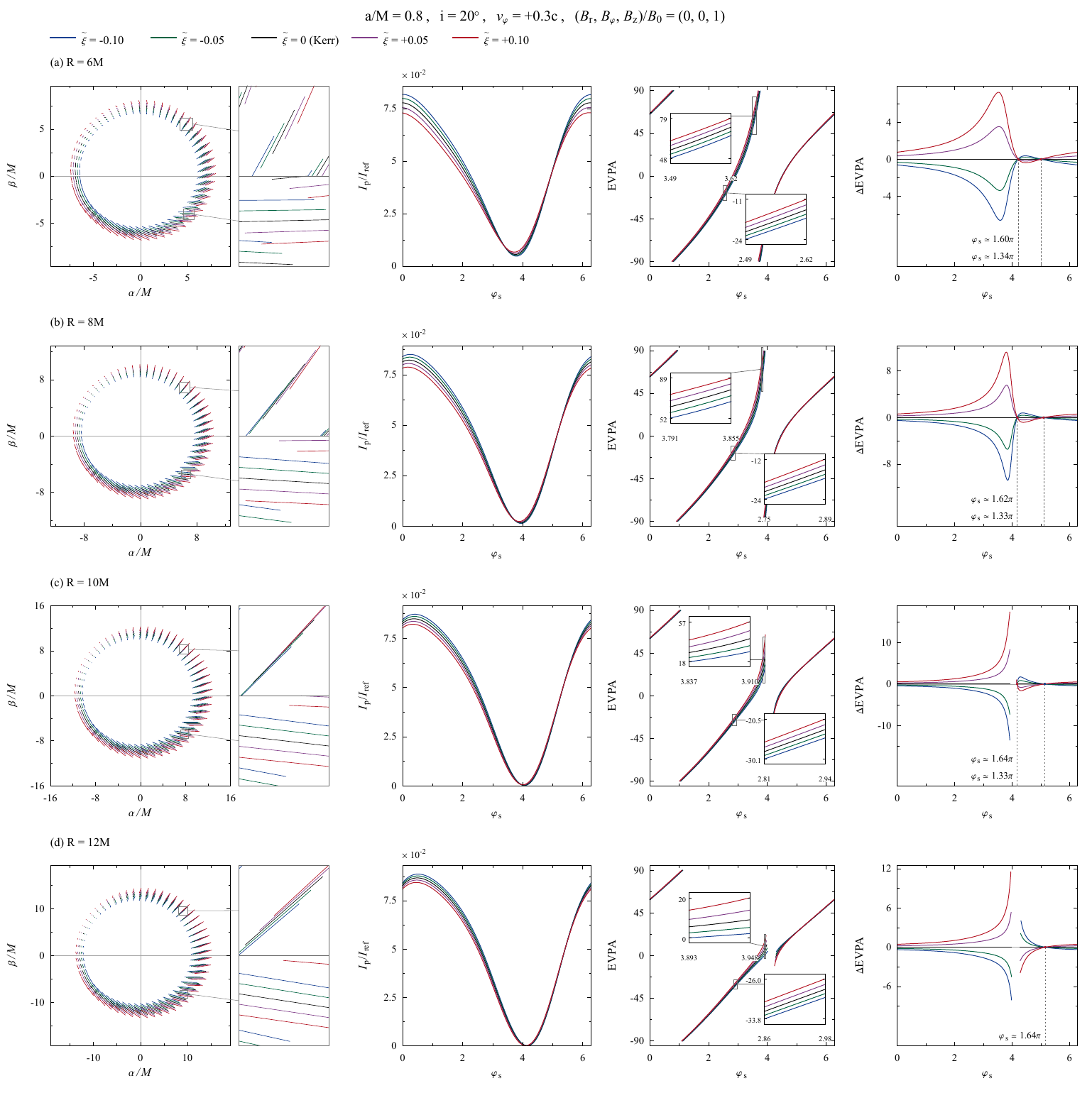}
	\caption{Emission-radius comparison for a vertical field and prograde flow (\(v_\phi=+0.3c\)) at \(a/M=0.8\) and \(i=20^\circ\). Rows (a)--(d) correspond to \(R/M=6,8,10,\) and \(12\).The angle difference is plotted only where the polarized intensity in each model exceeds \(1\%\) of its own peak.}
    \label{ER1}
\end{figure*}

\begin{figure*}[!htbp]
	\centering
	\includegraphics[width=\linewidth]{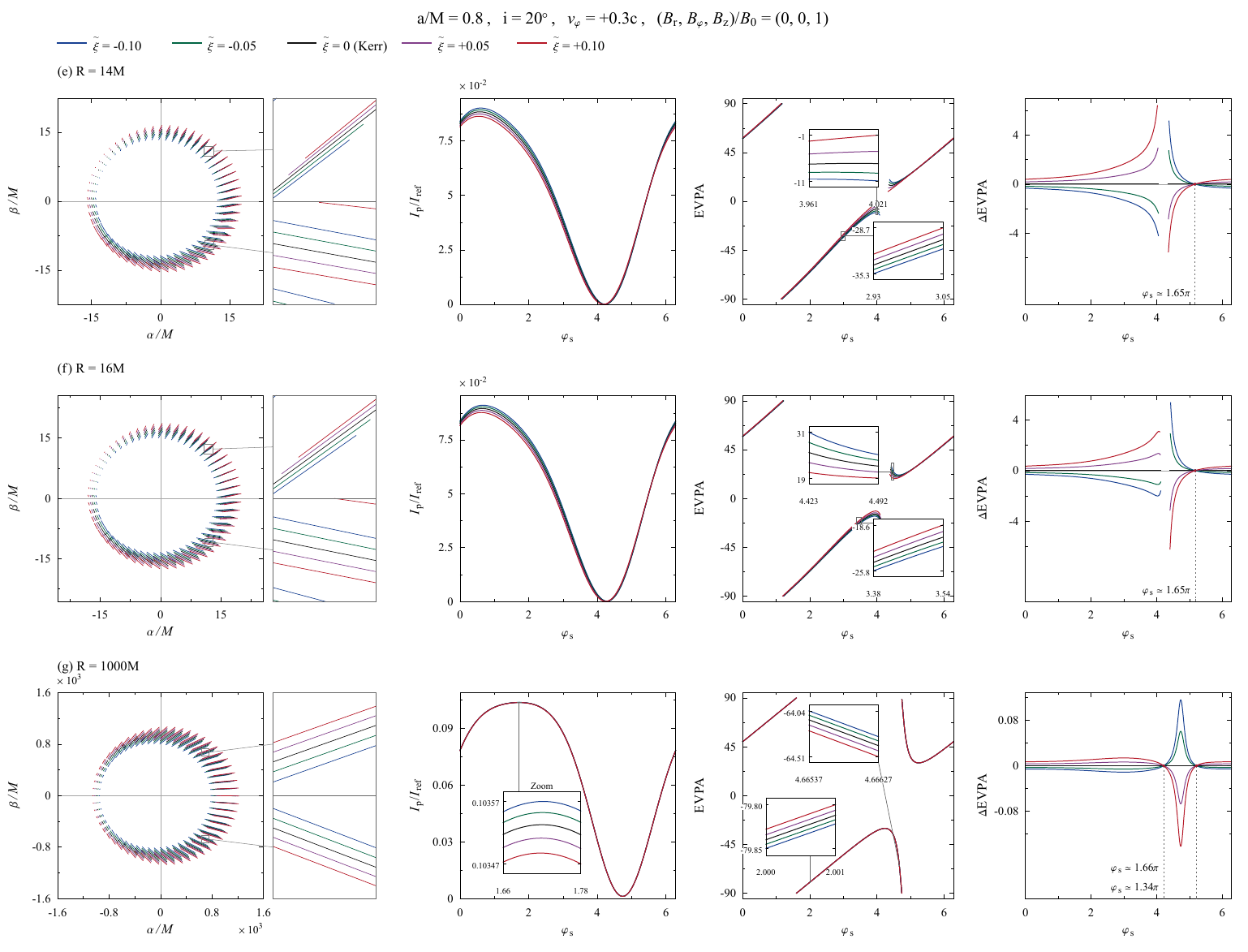}
	\caption{Continuation of the emission-radius comparison for a vertical field at \(a/M=0.8\), \(i=20^\circ\), and \(v_\phi=+0.3c\). Rows (e)--(g) show \(R/M=14,16,\) and \(1000\). }
    \label{ER2}
\end{figure*}

Fig.~\ref{spin}, the deformation changes
the polarized intensity and polarization direction. As \(a/M\) increases
from \(0\) to \(0.8\), the overall polarization pattern remains similar,
while the local \(\lvert\Delta\mathrm{EVPA}\rvert\) increases slightly.

In Figure~\ref{magnetic},
 the polarization bars are approximately tangential to the ring for a purely radial field and nearly radial for a purely toroidal field.
 The two mixed fields produce different bar orientations and bright-arc locations.
 Among the five configurations,
 the purely vertical field yields the lowest polarized intensity but the largest local
 $\lvert\Delta\mathrm{EVPA}\rvert$.

For the purely vertical field in Fig.~\ref{inclination}, the projected ring becomes
flatter and the peak polarized intensity rises as the inclination
increases from \(20^\circ\) to \(60^\circ\) and \(80^\circ\).
Over the same sequence, the maximum
\(\lvert\Delta\mathrm{EVPA}\rvert\) decreases.

Figure~\ref{fluid} shows a nonzero \(\lvert\Delta\mathrm{EVPA}\rvert\) even when the fluid
is at rest in the Z-frame. Reversing the azimuthal flow shifts both
the bright arc and the source azimuth of the dominant angular
deviation. For radial inflow, the intensity minimum is shallower
and the peak \(\lvert\Delta\mathrm{EVPA}\rvert\) is smaller than
for either azimuthal flow.

Figures~\ref{ER1} and~\ref{ER2} show how the polarization profiles vary with emission
radius. Among the seven radii shown, the largest local
\(\lvert\Delta\mathrm{EVPA}\rvert\) occurs at \(R=10M\), near a
minimum of the polarized intensity. At \(R=1000M\), the intensity
and EVPA profiles for the different deformations nearly coincide.

\subsection{Parameter dependence of the \(Q\)--\(U\) trajectories}
To further demonstrate the correlation between polarization intensity and polarization direction, we investigate the trajectory of the emission ring in the Stokes $Q-U$ plane.

\begin{figure*}[!htbp]
	\centering
	\includegraphics[width=\linewidth]{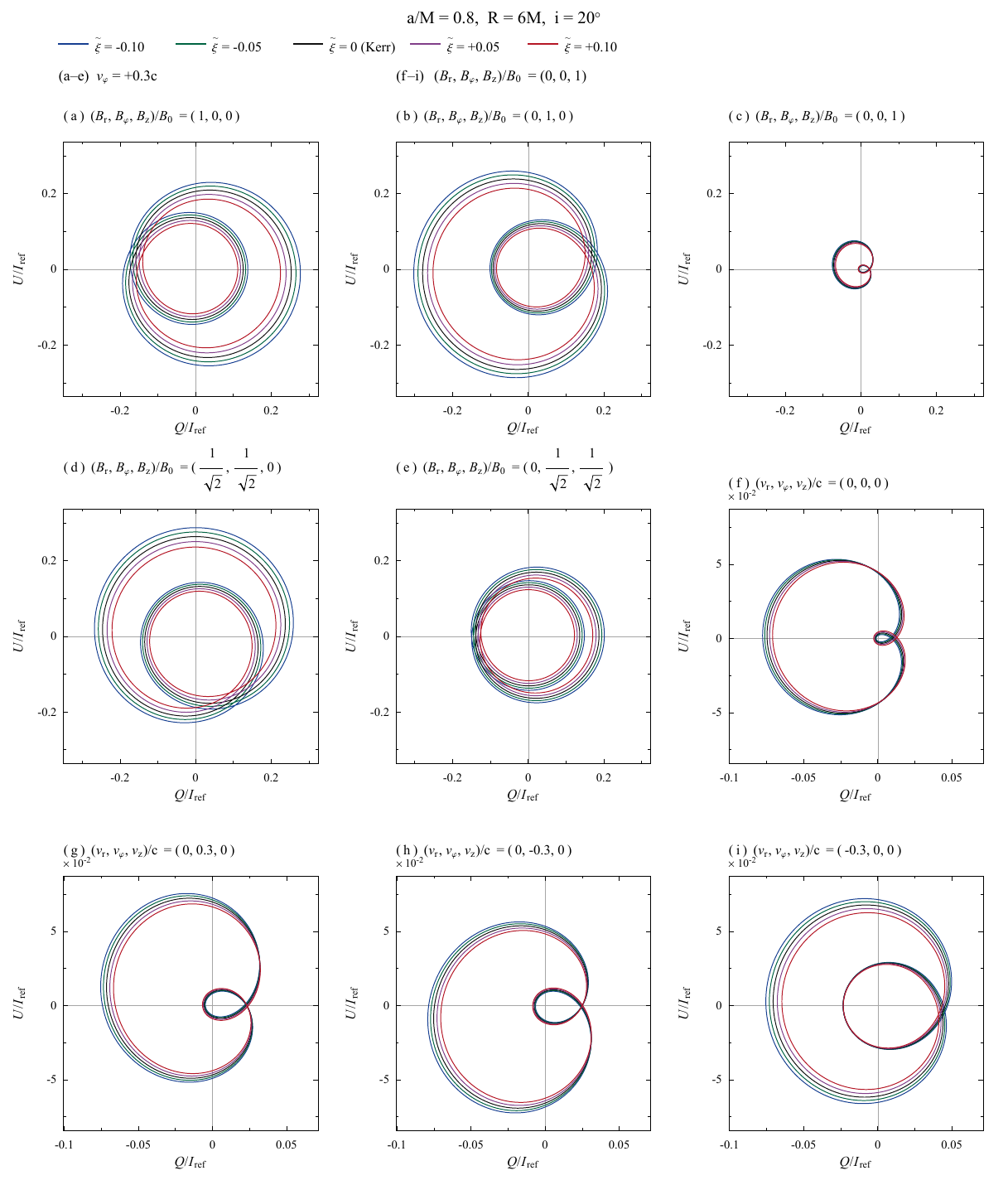}
	\caption{Magnetic field and flow comparisons in the normalized Stokes plane at \(a/M=0.8\), \(R=6M\), and \(i=20^\circ\). Each curve follows \((Q/I_{\rm ref},U/I_{\rm ref})\) once around the emitting ring. Panels (a)--(e) vary the field at \(v_\phi=+0.3c\), while panels (f)--(i) vary the flow in a vertical field. Panels (c) and (g) show the same configuration on different axis scales.}
    \label{mag_fu_QU}
\end{figure*}

\begin{figure*}[!htbp]
	\centering
	\includegraphics[width=\linewidth]{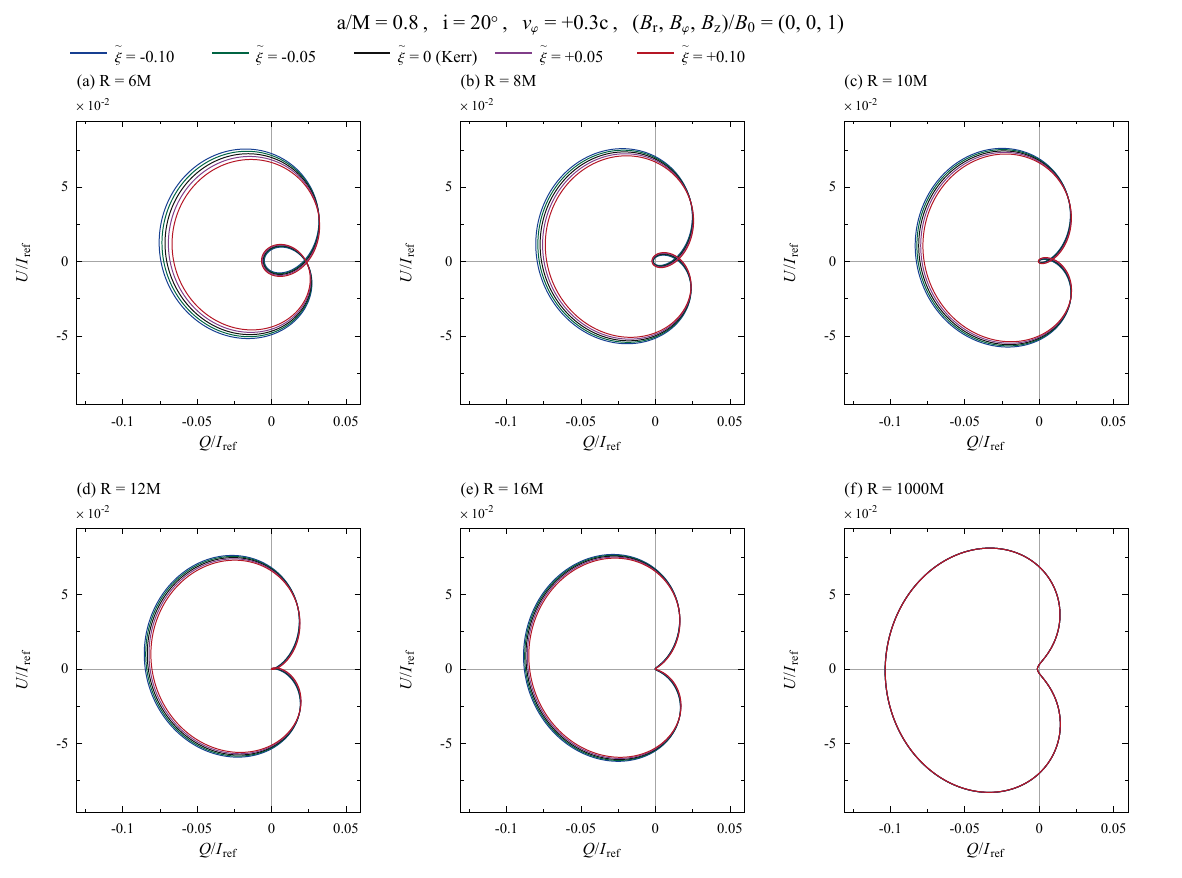}
	\caption{Radius dependence of the normalized Stokes trajectories for a vertical field, \(v_\phi=+0.3c\), \(a/M=0.8\), and \(i=20^\circ\). Panels (a)--(f) show \(R/M=6,8,10,12,16,\) and \(1000\) on identical axes. }
    \label{ERQU}
\end{figure*}

Panels (a)--(e) of Fig.~\ref{mag_fu_QU} compare the five magnetic
field configurations. The radial and toroidal fields both produce
broad trajectories with distinct inner and outer loops. The vertical
field gives the most compact trajectory, consistent with its lower
polarized intensity, while the two mixed fields differ in the relative
sizes of their inner and outer loops. For all five configurations,
increasing $\tilde{\xi}$ reduces the maximum distance of the trajectory
from the origin.

The effect of fluid motion is shown in
Fig.~\ref{mag_fu_QU}(f)--(i) for a purely vertical field.
The inner loop is small when the fluid is at rest in the Z-frame.
Reversing the azimuthal flow shifts the inner loop and changes the
asymmetry of the outer loop. Radial inflow produces the largest inner
loop relative to the outer one, consistent with the shallower minimum
in polarized intensity seen in the corresponding ring profiles.

Figure~\ref{ERQU} shows the trajectories at six emission radii
on a common scale. The inner loop is clearly visible at $R=6M$,
shrinks at $R=8M$, and lies close to the origin at $R=10M$. At \(R=10M\), the pronounced local EVPA difference occurs on a weakly
emitting segment of the ring and therefore has little effect on the outer
contour. As the emission radius increases, the separation between the
trajectories for different values of \(\widetilde{\xi}\) becomes less
pronounced.

\section{Summary}
This work investigates the direct polarized images of synchrotron-emitting rings around Finslerian Kerr black holes. We numerically integrate the null geodesic and parallel transport
equations. For each configuration, we compare the polarized images in
the deformed spacetime with those in the Kerr limit at the same metric
mass parameter \(M\) and under the same emission and viewing conditions.
We examine how magnetic field structure, fluid motion, observer inclination,
and emission radius affect the response to deformation.

Within the parameter range studied, increasing the deformation
parameter enlarges the apparent ring and changes the polarized
intensity and EVPA around it. The magnetic field geometry and
fluid motion affect both the location of the bright arc and the
azimuthal distribution of the EVPA shift relative to Kerr. For a
purely vertical field, increasing the inclination raises the peak
polarized intensity while reducing the maximum absolute EVPA shift.
Among the emission radii examined, the largest local shift occurs
near an intensity minimum at $R=10M$. At $R=1000M$, the intensity
and EVPA profiles for the different deformations nearly coincide.

The $Q$--$U$ diagrams show how these changes in intensity and
polarization direction appear together. Since the distance from
the origin equals the polarized intensity, the faint arc at
$R=10M$ traces a segment close to the origin. Its large local
EVPA shift therefore has little effect on the outer contour of
the trajectory. At $R=1000M$, the trajectories for the different
deformations nearly overlap.

From an observational perspective, the largest local EVPA shift in our radius scan occurs on a faint arc at \(R=10M\), 
where the low polarized intensity makes the EVPA harder to measure accurately. This local deviation from the Kerr spacetime may therefore be difficult to distinguish with current EHT observations.
Nekouee found that variations in spin and deformation can produce analogous shifts in characteristic orbital radii. 
However, our comparisons keep the spin fixed, 
leaving it an open question whether polarized ring images can distinguish models with similar orbital radii but different combinations of spin and deformation.

\section*{Declaration of competing interest}
The authors declare that they have no known competing financial
interests or personal relationships that could have appeared to influence
the work reported in this paper.

\section*{Acknowledgements}

This work was supported by the National SKA Program of China
(Grants Nos. 2022SKA0110200 and 2022SKA0110203). This work was
also supported by Xiaofeng Yang’s ZHISHAN Distinguished Professor
startup funding of Henan University.

\section*{Data availability}

This work is a theoretical work thus no data are used.
\bibliography{apssamp}

\end{document}